\documentclass[12pt,preprint]{aastex}

\begin{document}

\title{Detection of a Fully-resolved Compton Shoulder of the Iron K$\alpha$
Line in the \textit{Chandra} X-ray Spectrum of GX~301$-$2}

\author{Shin Watanabe\altaffilmark{1,2}, Masao Sako\altaffilmark{3,4}, Manabu Ishida\altaffilmark{5}, Yoshitaka Ishisaki\altaffilmark{5}, Steve M. Kahn\altaffilmark{6}, Takayoshi Kohmura\altaffilmark{7}, Umeyo Morita\altaffilmark{5}, Fumiaki Nagase\altaffilmark{1}, Frederik Paerels\altaffilmark{8} \and Tadayuki Takahashi\altaffilmark{1,2}}

\altaffiltext{1}{Institute of Space and Astronautical Science, 3-1-1 Yoshinodai, Sagamihara, Kanagawa 229-8510, Japan; watanabe@astro.isas.ac.jp, nagase@astro.isas.ac.jp, takahasi@astro.isas.ac.jp}
\altaffiltext{2}{Department of Physics, University of Tokyo, 7-3-1 Hongo, Bunkyo, Tokyo 113-0033, Japan}
\altaffiltext{3}{Theoretical Astrophysics and Space Radiation Laboratory, Caltech MC 130-33, Pasadena, CA 91125, USA; masao@tapir.caltech.edu}
\altaffiltext{4}{Chandra Postdoctoral Fellow}
\altaffiltext{5}{Department of Physics, Tokyo Metropolitan University, 1-1 Minami-Osawa Hachioji, Tokyo 192-0397, Japan; ishida@phys.metro-u.ac.jp, ishisaki@phys.metro-u.ac.jp, umeyo@phys.metro-u.ac.jp}
\altaffiltext{6}{The Kavli Institute for Particle Astrophysics and Cosmology,
 Stanford University, Stanford, CA 94305, USA; skahn@slac.stanford.edu}
\altaffiltext{7}{Kogakuin University, 2665-1 Nakano-cho, Hachioji, Tokyo 192-0015, Japan; tkohmura@map.kougakuin.ac.jp}
\altaffiltext{8}{Columbia Astrophysics Laboratory, 550 West 120th St., New York, NY 10027, USA; frits@astro.columbia.edu}

\received{2003 August 12}
\revised{2003 September 9}
\accepted{2003 September 11}

\shorttitle{Compton shoulder of GX~301$-$2}
\shortauthors{Watanabe et al.}

\begin{abstract}
We report the detection of a fully-resolved, Compton-scattered emission 
line in the X-ray spectrum of the massive binary GX~301$-$2 obtained 
with the High Energy Transmission Grating Spectrometer onboard the 
\textit{Chandra} X-ray Observatory. The iron K$\alpha$ fluorescence 
line complex observed in this system consists of an intense narrow 
component centered at an energy of $E$~=~6.40~keV and a redward 
shoulder that extends down to $\sim$~6.24~keV, which corresponds to an 
energy shift of a Compton back-scattered iron K$\alpha$ photon. From 
detailed Monte Carlo simulations and comparisons with the observed 
spectra, we are able to directly constrain the physical properties of the 
scattering medium, including the electron temperature and column density, 
as well as an estimate for the metal abundance.

\end{abstract}

\section{Introduction}
When a high energy photon propagating through a low-temperature 
($<$~10$^5$~K) medium undergoes Compton scattering with the constituent 
electrons, the energy of the photon is modified in a way that depends on 
the scattering angle and the electron velocity distribution.  In the 
limit where the electrons are at rest, a fraction of the photon energy 
is transferred to the electron according to the Compton formula, 
\begin{equation}
  E_1 = \frac{E_0}{1 + \left( E_0/m_e c^2 \right)\left(1 - \cos \theta
  \right)},
\end{equation}
where $E_0$ is the energy of the incoming photon, $E_1$ is the energy of 
the outgoing photon, $\theta$ is the angle between the incoming and 
outgoing photons, and $m_e c^2$ is the electron rest-mass energy 
(=~511~keV). The maximum energy shift per scattering due to the electron 
recoil is, therefore, $\Delta E_\mathrm{max} = 2 E_0^2/(m_e c^2 + 2 E_0)$ 
for photons that are back-scattered ($\theta$~=~180$^\circ$). An X-ray 
emission line propagating through a medium with substantial Compton 
optical depth ($\tau_\mathrm{Compton}$~$>$~0.1) has a non-negligible 
probability of interacting with an electron, resulting in down-scattering 
of photons and, hence, producing a discernible ''Compton shoulder'' 
between $E_0$ and $E_0 -\Delta E_\mathrm{max}$.

X-ray emission lines at higher energies are ideal for studying the properties
of the Compton shoulders, since the Compton scattering opacity relative to
that of photoelectric absorption is larger for higher energy photons.  The
iron K$\alpha$ fluorescent line complex at $E_0$~=~6.40~keV, therefore, is
particularly promising, and can be produced over an extremely wide range in
column density, which makes it ideal for diagnosing the physical properties of
a cold medium irradiated by X-rays \citep{hatchett77, basko78, illarionov79,
george91, sunyaev96, matt02}.  The energy shift of an iron line photon due to
a single Compton scattering $\Delta E_\mathrm{max}$~=~156~eV (see eq.(1) ) is,
however, at best, comparable to the spectral resolving power of past X-ray
detectors in orbit.  Previous detections of the Compton shoulder are
unresolved (see, e.g., \citealt{molendi03}) and, in many cases, they are
statistically marginal as well \citep{iwasawa97, matt03}. The high spectral
resolution of 33~eV (FWHM) at 6.4~keV attained by the High Energy Transmission
Grating Spectrometer (HETGS) \citep{canizares00} onboard the \textit{Chandra}
X-ray Observatory \citep{weisskopf00} enables us, for the first time, to
resolve the Compton shoulder of the iron K$\alpha$ line (see, e.g.,
\citealt{kaspi02,bianchi02}).

The X-ray binary pulsar GX~301$-$2 is one of the most promising targets 
for studying the Compton shoulder associated with the 6.4~keV iron 
K$\alpha$ line. The system consists of an accreting magnetized neutron 
star in a highly eccentric orbit ($e$~=~0.46; \citealt{sato86, koh97}), embedded 
in the stellar wind from a B2 super-giant companion star \citep{bord76}. 
X-ray continuum photons from the neutron star ionize the K-shell 
electrons of iron in the wind, a fraction of which is followed by 
fluorescence emission of the 6.4~keV line. Previous observations of the 
system have shown an intense iron line with an equivalent width as large 
as $\sim$~1~keV \citep{white84, leahy89, tashiro91, endo02}. Given the 
large mass-loss rate of the companion \citep{parkes80}, a substantial 
fraction of the 6.4~keV photons can subsequently undergo Compton 
scattering prior to escaping the wind.

In this \textit{Letter}, we present spectral analysis results of the 
properties of the iron K$\alpha$ line and its Compton shoulder observed 
in GX~301$-$2 with the \textit{Chandra} HETGS.  As we show, the detection 
of the fully-resolved Compton shoulder provides a unique opportunity to 
investigate in detail the physical state and geometrical distribution of 
cold ($<$~10$^5$~K) material surrounding an X-ray source. In \S~2, we 
describe the observation and the procedures adopted for data reduction. 
Some basic spectral measurements and observed quantities are also 
discussed. In \S~3, we describe our Monte Carlo simulator used for the 
spectral analyses.  Finally in \S~4, we present our measurements and 
discuss the implications of the results.

\section{Observation and Data Reduction}
\textit{Chandra} observed GX~301$-$2 at three different orbital phases: 
(1) $\phi$~=~0.167--0.179, (2) $\phi$~=~0.480--0.497, and 
(3) $\phi$~=~0.970--0.982, hereafter referred to as intermediate (IM), 
near-apastron (NA), and pre-periastron (PP) phases \citep{pravdo95}, 
respectively.  We focus primarily on the spectrum of the PP-phase, during 
which the X-ray luminosity, absorbing column density, and the iron line 
equivalent width are the highest \citep{endo02}. The observation of the 
PP-phase was performed on 2002 February 3 12:34:10 UT and continued for
$\sim$~40~ksec.  All of the data were processed using CIAO v2.2.1. 
Since the zeroth order image was severely piled-up, the centroid was 
determined by finding the intersection of the streak events and the 
dispersed events.  We apply a standard spatial filter and an order 
sorting mask to extract only the first order events. The exposure map 
and effective area were calculated based on the extraction region 
described above.  Since we are interested mainly in the properties of 
the iron line, we use data only from the High-Energy Grating (HEG) for 
the analysis, since it has a higher resolving power 
($\Delta E_\mathrm{FWHM}$~=~33~eV at 6.4~keV) and larger effective area 
(20~cm$^2$) in the iron K line region.

The X-ray light curve in the 1.0--10.0~keV band extracted from the HEG 
events is shown in the top panel of Fig.~1.  An X-ray outburst is seen 
in the first half of the observation.  Hence, we have chosen to divide 
the data into the first and second halves as indicated by the vertical 
dotted line, and have extracted spectra from each data segment 
separately. The observed 2--10~keV fluxes are 
13.2~$\times$~10$^{-10}$~erg~cm$^{-2}$~s$^{-1}$ and 
9.0~$\times$~10$^{-10}$~erg~cm$^{-2}$~s$^{-1}$ during the first and 
second halves, respectively.  A blow-up of the iron K$\alpha$ line 
spectra are shown in the bottom panels of Fig.~1. In addition to the 
intense iron K$\alpha$ line at 6.4~keV, a shoulder extending towards 
the low energy side down to $\sim$~6.24~keV is clearly seen in the data. 
The equivalent widths of the iron K$\alpha$ lines including the shoulders 
are 643~$\pm$~20~eV and 486~$\pm$~18~eV for the first and second halves, 
respectively.  The width of the shoulder ($\Delta E$~$\sim$~160~eV) 
precisely matches what is predicted by eq.~(1), which strongly indicates 
that the feature is formed primarily through single Compton scattering of 
the iron K$\alpha$ photons.  Changes in the shape of the Compton shoulder 
as well as in the shoulder flux relative to the un-scattered line flux, 
can be seen between the spectra of the first and second halves.

\section{Monte Carlo Simulation}
The Compton shoulder can be used to infer various physical parameters 
that characterize the scattering medium.  The flux ratio of the shoulder 
to the line is determined by the metal abundance and the optical thickness 
of the scattering cloud.  Its energy distribution, on the other hand, is 
sensitive to the temperature and the geometrical distribution of the 
scattering electrons. A discernible change in the profile shown in Fig.~1 
implies that these physical parameters are variable between the first and 
second halves of the observation.

In order to obtain some quantitative information from the spectra, we have
constructed a Monte Carlo simulator to compute the emergent spectrum from an
X-ray source surrounded by a cloud.  We assume a spherical distribution of
material motivated by the observed correlation between the iron emission line
strength and its corresponding edge depth during the IM, NA, and PP phases.
This assumption also significantly simplifies the simulation.  The cloud
consists of H, He and astrophysically abundant metals (C, N, O, Ne, Na, Mg,
Al, Si, S, Cl, Ar, Ca, Cr, Fe and Ni), and all of the metal abundances
(Z~$>$~2) are allowed to vary together relative to the cosmic values of
\citet{feldman92}.  We account for photoelectric absorption and subsequent
fluorescent emission, as well as Compton scattering by free electron. The
angular-dependence of the Compton scattering cross section is fully accounted
for and the electrons are assumed to have a Maxwellian energy
distribution. The photons may suffer multiple interactions and are traced
until they completely escape the cloud.  The energy distribution of the
emergent photons are then histogrammed to produce a spectrum.

Fig.~2 shows some of the results from the simulations for the iron line and
its Compton shoulder with varying hydrogen column density ($N_\mathrm{H}$) and
electron temperature ($kT_\mathrm{e}$). The original iron K$\alpha_1$ and
K$\alpha_2$ photons are assumed to be distributed according to the K-shell
photoionization rate at each radius from the central continuum source. The
distribution of the original iron K photons is also calculated with the
simulator, in which a power-law X-ray source with a photon index of 1.0 is
assumed.  In the upper panels of Fig.~2, one can see an increase in the
scattered flux relative to the narrow line flux as $N_\mathrm{H}$ is
increased. The lower panels of Fig.~2 show the temperature dependence of the
shape of the Compton shoulder.  More smearing is seen at higher
$kT_\mathrm{e}$.  Photons between 6.24~keV and 6.40~keV are due primarily to
single-scattered photons, and the component below 6.24~keV result from
multiple-scattering, which are important even at moderate optical depths.

\section{Discussion}
We performed spectral fits based on the simulation for the iron line and 
the Compton shoulder.  A fits file of "table models" were generated from 
the results of the simulations, which were then incorporated into XSPEC 
v11.2. The parameters of the model are $N_\mathrm{H}$, $kT_\mathrm{e}$, 
and the metal abundances.  The radial dependence of the intrinsic 
K$\alpha$ line emissivity is fixed to what one expects from a photon 
index of $\Gamma = 1.0$.  We have confirmed that this assumption produces 
at most a 5~\% error in the emergent line profile for $\Gamma$ between 
0.0 and 2.0.

\subsection{Metal Abundance}
We first attempt to determine the metal abundances from the total PP-phase 
spectrum. In doing this, we note that two independent constraints between 
the abundance and the hydrogen column density are available, as described 
below. The first one arises from the observed intensity ratio of the 
shoulder to the narrow line.  For a given ratio, an increase in column 
density must be accompanied by an increase in abundance and, hence, the 
two parameters are correlated as shown in Fig.~3.  The other constraint 
originates from the observed equivalent width of the line. In this case, 
the abundance is anti-correlated with the column density. From these two 
constraints (see Fig.~3), the metal abundance is determined to be 
0.65--0.82 times of the cosmic value \citep{feldman92}, assuming 
$\Gamma$~=~1.0. The slope, however, is not well-measured due to the 
limited bandpass of the Chandra gratings.  If we allow a range in 
$\Gamma$ of 1.0--1.5, which has been seen in previous hard X-ray 
observations \citep{pravdo95, orlandini00}, the metal abundance is 
determined to be 0.65--0.90 times of the cosmic. Note that this is 
consistent with that of typical OB-stars \citep{daflon01}.

\subsection{Hydrogen Column Density and Electron Temperature}
Adopting the metal abundance of 0.75 cosmic derived in \S~4.1, we find 
the values for $N_\mathrm{H}$.  The derived values are listed in 
Table~1, and the models are shown by the lines superimposed on 
the data in Fig.~1.  The difference in the observed Compton profile can be 
described by a change in the column density, which results in a variation 
in the number-of-scatterings distribution even at these moderate optical 
depths. We have also obtained upper-limits (90~\% confidence levels) to 
$kT_\mathrm{e}$ of $<$~3.4~eV and $<$~0.6~eV for the first- and 
second-half spectra, respectively.  Interestingly, the derived electron 
temperatures are consistent with that expected for the stellar wind being 
cooled via adiabatic expansion from the surface of the companion star with 
a photospheric temperature of $\sim$~2~eV \citep{bord76}. Interestingly, 
the column density as inferred from the Compton profile, in fact, 
reproduces the spectrum in the entire HEG bandpass for a power-law photon 
index of 1.0.  Fig.~4 shows the observed spectra overlaid with the 
simulated spectra.  Reduced chi-squared values of 1.30 and 0.95 were 
obtained for the first and second halves spectra, respectively. 
Though some residual flux still remain, the simulations, which are based on
parameters derived from the line and Compton shoulder, provide fairly good
descriptions of the broadband data.
Assuming a distance of 1.8~kpc \citep{parkes80}, the absorption-corrected
X-ray luminosities in the 2--10~keV are 
3.5~$\times$~10$^{36}$~erg~s$^{-1}$ and 
1.6~$\times$~10$^{36}$~erg~s$^{-1}$ for the first and second halves, 
respectively.

\subsection{Future Prospects}
We have demonstrated through our quantitative analysis of the 
\textit{Chandra} HETGS data of GX~301$-$2 that the Compton shoulder 
provides a sensitive diagnostic for the physical conditions of cold 
material irradiated by X-ray photons. The substantial improvement in the 
spectral resolution (6~eV at 6~keV) and the effective area 
($\sim$~300~cm$^{2}$ at 6~keV) accessible with the X-ray Spectrometer 
(XRS) onboard the Astro-E2 observatory, which is planned to be deployed 
in 2005, will allow us to study the Compton shoulder profile in 
various classes of X-ray sources, such as a molecular cloud irradiated by 
a proto-star, an accretion disk swirling around a compact object, and a 
dusty torus surrounding an active galactic nucleus.  More detailed 
geometrical information can be obtained by studying the energy 
distribution and, hence, the scattering angle distribution in the 
circumsource medium.  Compton scattering by electrons in bound systems 
produces spectral signatures that are distinct from those produced by 
free electrons \citep{sunyaev96, vainshtein98, sunyaev99}, which may be
detectable with future X-ray spectrometers.

\acknowledgements {\small SW} is supported by research
fellowships of the Japan Society for the Promotion of Science for young
Scientists. {\small MS} was supported by NASA through the 
Chandra Fellowship Program.

\clearpage
\begin{deluxetable}{rcccc}
\tablecolumns{5}
\tablewidth{0pt}
\tablecaption{Derived Parameters from Spectral fits of the Compton shoulder \label{tbl:fits2}}
\tablehead{
  & $N_\mathrm{H}$ ($10^{23}$~cm$^{-2}$) & $\tau_\mathrm{Compton}$ & $kT_\mathrm{e}$ (eV) (upper limit) & $\chi^2$/ d.o.f.\tablenotemark{a}
}
\startdata
First  & 12.0$^{+3.5}_{-1.3}$ & 0.96$^{+0.28}_{-0.10}$ & 0.5 ($<$~3.4) & 65.6 / 71 \\
Second &  8.5$^{+2.3}_{-1.4}$ & 0.68$^{+0.18}_{-0.11}$ & 0.0 ($<$~0.6) & 82.4 / 71 \\
\enddata

\tablecomments{Errors and upper limits designate 90~\% confidence levels.}
\tablenotetext{a}{Only the 6.0 -- 6.6~keV region has been used in the spectral
fit.}

\end{deluxetable}
\clearpage

\begin{figure}
\epsscale{0.75}
\plotone{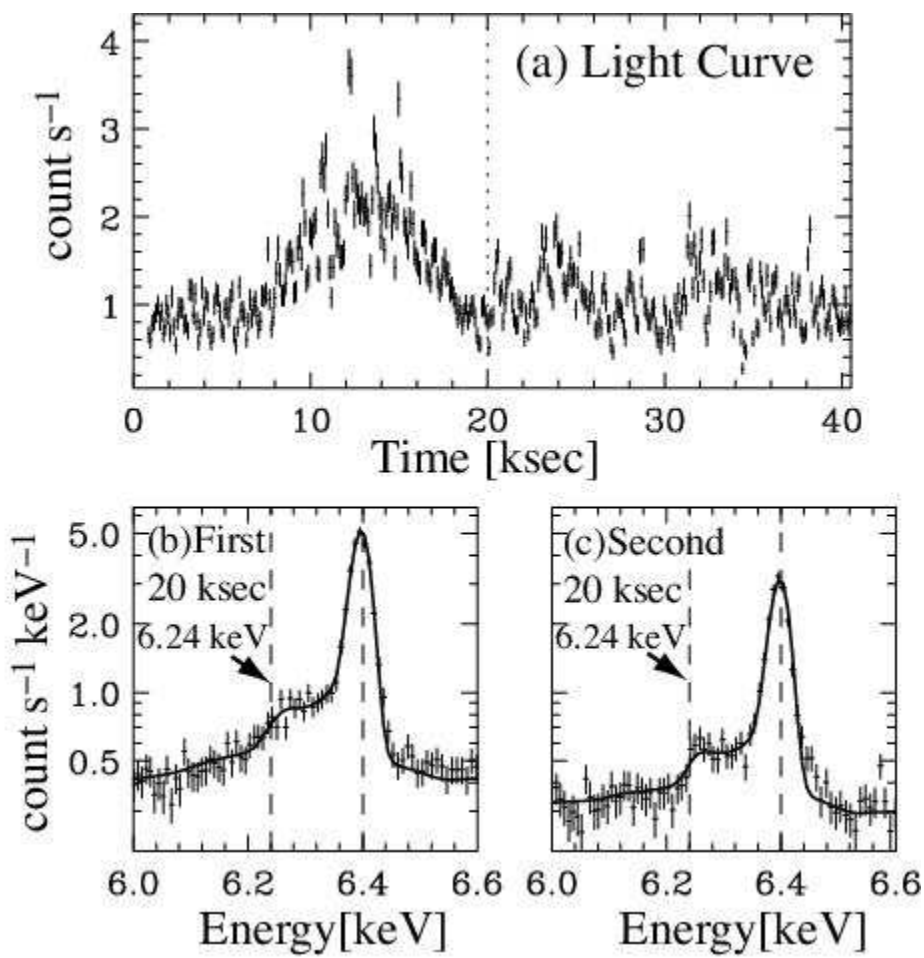}
\figcaption{(a): The light curve of GX~301$-$2 in the 
pre-periastron (PP) phase observed with the \textit{Chandra} HETGS in the 
band 1--10~keV ($\pm$~1 order of HEG). (b),(c): The spectra of the iron 
K$\alpha$ region (6.0--6.6~keV) for the first and the second halves of the 
observation as defined by the dotted vertical line shown in panel (a). 
Each spectrum shows a clear drop in flux near 6.24~keV. Superimposed on 
the data in lines are the best-fit Monte Carlo models (see text and 
Fig.~2 ). \label{fig:obs}}
\end{figure}

\begin{figure*}
\epsscale{0.8}
\plotone{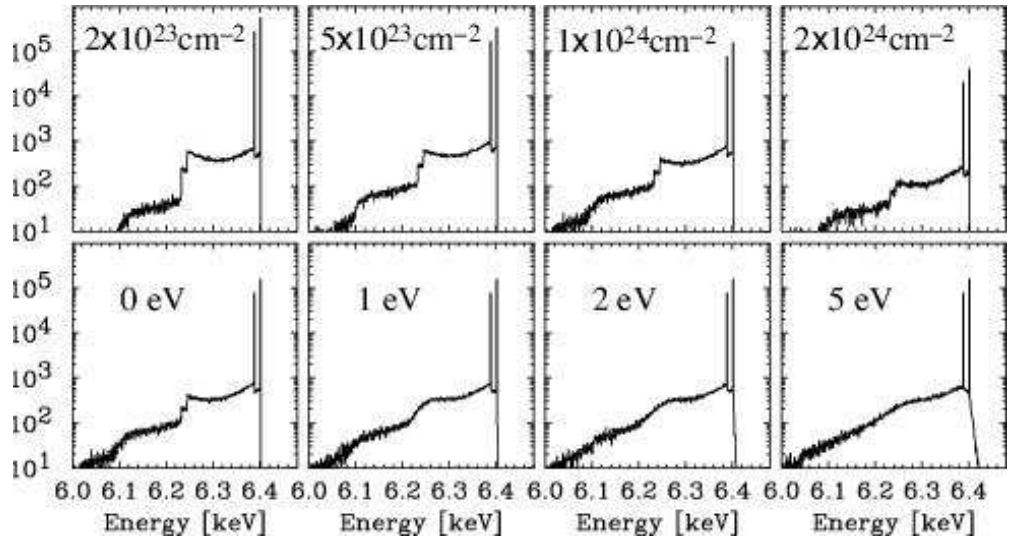}
\figcaption{Dependence of the iron K$\alpha$ line profile on the hydrogen 
column density ($N_\mathrm{H}$) and the electron temperature 
($kT_\mathrm{e}$). The upper panels show the variation of the iron 
K$\alpha$ line and its shoulder as a function $N_\mathrm{H}$ for a fixed 
$kT_\mathrm{e}$ at 0~eV. The lower panels show the variation as a function 
of $kT_\mathrm{e}$ between 0~eV and 5~eV for a fixed $N_\mathrm{H}$ at 
1~$\times$~10$^{24}$~cm$^{-2}$. In these simulations, the metal abundances 
were assumed to be 0.75 times cosmic \citep{feldman92}. \label{fig:sim}}
\end{figure*}

\begin{figure}
\epsscale{0.75}
\plotone{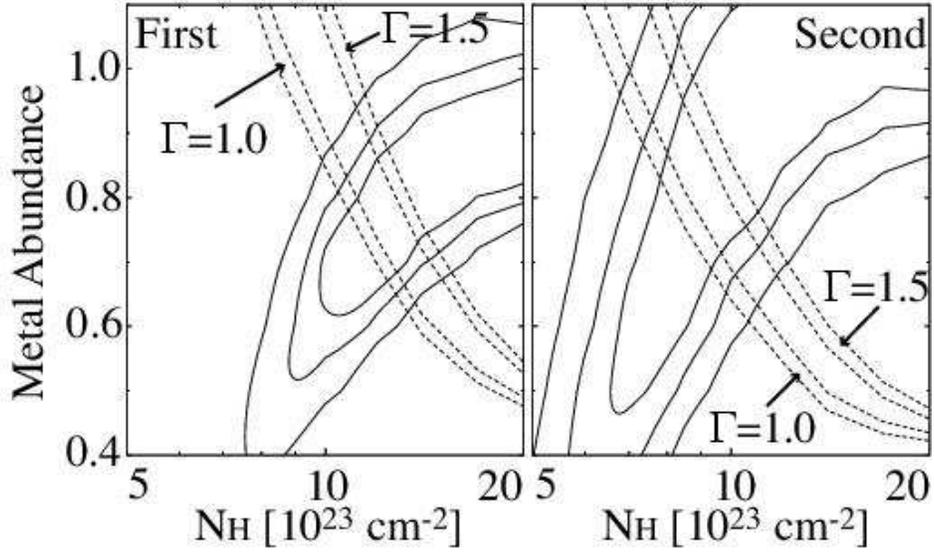}
\figcaption{Confidence contour for $N_\mathrm{H}$ vs. the metal abundance 
during the first half (left) and the second half (right). Each contour 
represents 68~\%, 90~\% and 99~\% confidence level. The region between each 
pair of dashed lines are the values allowed by the observed line equivalent 
widths of 643~$\pm$~20~eV (left) and 486~$\pm$~18~eV (right) for two 
assumed values for the photon index ($\Gamma$~=~1.0 and $\Gamma$~=~1.5). 
The metal abundance is determined to be 0.65--0.90 (90~\% confidence range) 
times the cosmic value. \label{fig:contour}}
\end{figure}

\begin{figure}
\epsscale{0.75}
\plotone{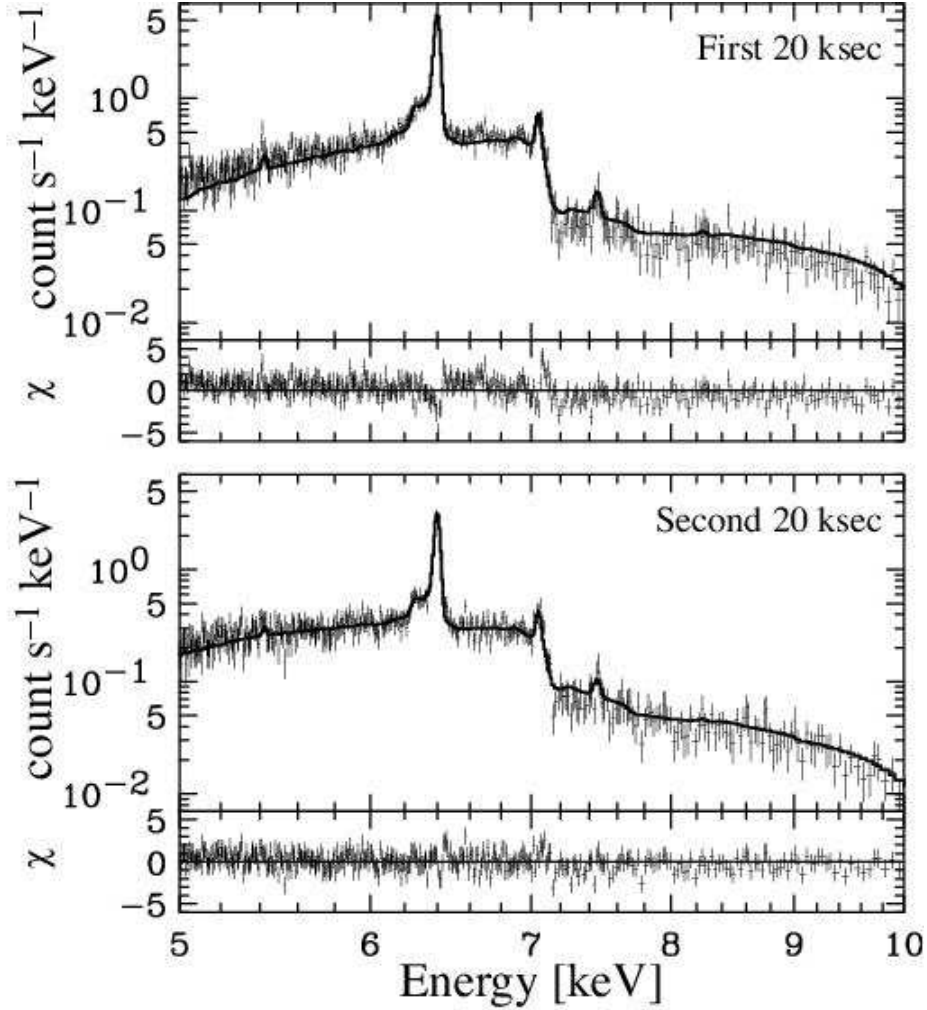}
\figcaption{The spectra of GX~301$-$2 from 5~keV to 10~keV for the first 
and second halves of the PP-phase observation. The solid lines show the 
Monte Carlo models inferred from the data. In addition to the iron 
K$\alpha$ emission line profile, the continuum shape is also successfully 
reproduced with the parameters inferred from the Compton shoulder profile. 
The values for $N_\mathrm{H}$ are 12.0~$\times$~10$^{23}$~cm$^{-2}$ and 
8.5~$\times$~10$^{23}$~cm$^{-2}$ for the first and second halves, 
respectively. A photon power-law index of 1.0 is assumed.\label{fig:all}}
\end{figure}
\end{document}